\documentstyle[preprint,aps]{revtex}
\begin{document}
\preprint{OITS-538}
\draft
\title{Hadronic Penguin B Decays In The Standard\\
And The Two-Higgs-Doublet Models
\footnote{Work supported in part by the Department of Energy Grant No.
DE-FG06-85ER40224.}}
\author{N.G. Deshpande and
Xiao-Gang He}
\address{Institute of Theoretical Science\\
University of Oregon\\
Eugene, OR 97403-5203, USA}
\date{March, 1994; Revised May, 1994}
\maketitle
\begin{abstract}
We study in next-to-leading order QCD hadronic penguin $B$ decays
in the Standard and two-Higgs-doublet models. Although the gluonic penguin
dominates, we find the electroweak contribution non-negligible.
In the Standard Model, the branching ratio for $B \rightarrow X_s
\phi$ is predicted to be in the range $(0.6\sim 2)\times 10^{-4}$.
The ranges of branching ratios for $B\rightarrow K \phi$,
$B\rightarrow K^*\phi$, and $B_s\rightarrow \phi\phi$ are $(0.4\sim
2)\times 10^{-5}$, $(0.2\sim 1)\times 10^{-5}$, and
$(0.15\sim 0.5)\times 10^{-5}$, respectively.
The contribution from the charged Higgs boson in two Higgs doublet models
depend on $cot\beta$, and can be as large as 40\%.
\end{abstract}
\newpage

Rare $B$ decays, particularly pure penguin decays, have been subject of
considerable theoretical and experimental
interest recently\cite{bdecay}. The photonic
penguin induced process $B\rightarrow K^* \gamma$ has been observed
by CLEO collaboration\cite{cleo} and is consistent with the
Standard Model (SM) prediction\cite{desh1}. The gluonic penguin induced $B$
decays are expected
to be observed very soon. A large number of gluonic penguin induced $B$ decay
channels were
studied in Ref.\cite{desh2} using  $\Delta B = 1$ effective Hamiltonian
$H_{\Delta B=1}$ in the lowest nonvanishing order. In Ref.\cite{gatto1} the
next-to-leading order QCD corrected pure gluonic penguin $H_{\Delta B=1}$
was used with top quark mass $m_t$ fixed at
150 GeV. In this paper we study the next-to-leading order QCD corrected
Hamiltonian $H_{\Delta B = 1}$ in the SM and in two Higgs doublet models,
taking particular care to {\em include the full electroweak contributions} and
find the dependence on $m_t$ and $\alpha_s$. Using this
Hamiltonian we study the cleanest signature of hadronic penguin
processes: $B\rightarrow X_s \phi$, $B\rightarrow K\phi (K^*\phi)$, and
$B_s\rightarrow \phi\phi$. The process $B\rightarrow X_s\phi$ is particularly
recommended because it is free from form factor uncertainties. We find not only
that the QCD correction in next-to-leading order are large, but also inclusion
of the full electroweak contributions have significant effect on the branching
ratio which could reduce the pure gluonic penguin contribution by 30\% at the
upper range of allowed top quark mass. Our results which have been derived
independently, agree with Ref.\cite{fleischer} where only the SM is considered.

\noindent
{\bf $\Delta B = 1$ gluonic penguin Hamiltonian}

The QCD corrected $H_{\Delta B = 1}$ relevant to us
can be written as follows\cite{laut}:
\begin{eqnarray}
H_{\Delta B=1} = {G_F\over \sqrt{2}}[V_{ub}V^*_{us}(c_1O^u_1 + c_2 O^u_2)
+V_{cb}V^*_{cs}(c_1O^c_1+c_2O^c_2) - V_{tb}V^*_{ts}\sum c_iO_i] +H.C.\;,
\end{eqnarray}
where the Wilson coefficients (WCs) $c_i$ are defined at the scale of $\mu
\approx
m_b$; and $O_i$ are defined as
\begin{eqnarray}
O^q_1 = \bar s_\alpha \gamma_\mu(1-\gamma_5)b_\beta\bar
q_\beta\gamma^\mu(1-\gamma_5)q_\alpha\;&,&\;\;\;\;
O^q_2 = \bar s \gamma_\mu(1-\gamma_5)b\bar
q\gamma^\mu(1-\gamma_5)q\;,\nonumber\\
O_3 = \bar s \gamma_\mu(1-\gamma_5)b \sum_{q'}
\bar q' \gamma_\mu(1-\gamma_5) q'\;&,&\;\;\;\;
Q_4 = \bar s_\alpha \gamma_\mu(1-\gamma_5)b_\beta \sum_{q'}
\bar q'_\beta \gamma_\mu(1-\gamma_5) q'_\alpha\;,\nonumber\\
O_5 =\bar s \gamma_\mu(1-\gamma_5)b \sum_{q'} \bar q'
\gamma^\mu(1+\gamma_5)q'\;&,&\;\;\;\;
Q_6 = \bar s_\alpha \gamma_\mu(1-\gamma_5)b_\beta \sum_{q'}
\bar q'_\beta \gamma_\mu(1+\gamma_5) q'_\alpha\;,\\
O_7 ={3\over 2}\bar s \gamma_\mu(1-\gamma_5)b \sum_{q'} e_{q'}\bar q'
\gamma^\mu(1+\gamma_5)q'\;&,&\;\;
Q_8 = {3\over 2}\bar s_\alpha \gamma_\mu(1-\gamma_5)b_\beta \sum_{q'}
e_{q'}\bar q'_\beta \gamma_mu(1+\gamma_5) q'_\alpha\;,\nonumber\\
O_9 ={3\over 2}\bar s \gamma_\mu(1-\gamma_5)b \sum_{q'} e_{q'}\bar q'
\gamma^\mu(1-\gamma_5)q'\;&,&\;\;
Q_{10} = {3\over 2}\bar s_\alpha \gamma_\mu(1-\gamma_5)b_\beta \sum_{q'}
e_{q'}\bar q'_\beta \gamma_mu(1-\gamma_5) q'_\alpha\;.\nonumber
\end{eqnarray}

The WCs $c_i$ are obtained by solving
the renormalization group equation
\begin{eqnarray}
(\mu {\partial \over \partial \mu} + \beta (g) {\partial \over \partial g})
{\bf C}(m_W^2/\mu^2, g^2) = \hat \gamma^T(g^2) {\bf C}(m_W^2/\mu^2, g^2)\;.
\end{eqnarray}
Here ${\bf C}$ is the column vector $(c_i)$, and
\begin{eqnarray}
\beta (g) &=& -(11-{2\over 3}n_f){g^3\over 16\pi^2} - (102-{38\over 3}n_f)
{g^5\over (16\pi^2)^2} + ...\;,\nonumber\\
\hat \gamma(g^2)&=&\gamma^{(0)}_s {\alpha_s \over 4\pi} +
\gamma^{(1)}_s{\alpha^2_s\over(4\pi)^2}+ \gamma_e^{(0)} {\alpha_{em}\over 4\pi}
+\gamma^{(1)}_{se}{\alpha_{em}\alpha_s\over (4\pi)^2}+... \;,
\end{eqnarray}
where $n_f$ is the number of active quark flavours.

The anomalous-dimension matrix $\gamma^{(0)}_s$ and the first term in
$\beta(g)$
determine the leading log QCD corrections\cite{bruce}.  The rest of the
terms contain information about the leading QED  and next-to-leading order QCD
corrections. The full $10\times 10$ matrices for $\gamma_i$ are given
in Ref.\cite{laut}. The matching conditions
of the Wilson coefficients at $m_W$ for the next-to-leading order corrections
will be different from the leading order ones. One needs to
include one loop current-current corrections for $c_{1,2}$ at $m_W$. The full
results for the initial conditions can be found in \cite{laut}.

The WCs obtained above depend on the  renormalization regularization scheme
(RS) used. In our calculation we used the naive dimentional regularization
scheme.
The physical quantities, of course,  should not depend on RS provided one
handels the hadronic matrix elements correctly. In
practice, many of the hadronic matrix elements can only be calculated using
factorization method. In our later calculation we will also use this
approxmation. Since this approximation does not carry information about the
RS dependence, it is better for us to use WCs,
${\bf \bar C}(\mu) = ({\bf 1}+ {\bf \hat r_s}^T\alpha_s(\mu)
/4\pi +{\bf \hat r_e}^T\alpha_{em}(\mu)/4\pi){\bf C}(\mu)$, which are
RS independent\cite{buras}.
Here the matrices ${\bf \hat r}_{s,e}$ are obtained from one-loop matching
conditions. The $6\times6$ ${\bf \hat r}_s$ matrix for the pure gluonic penguin
operators has been given in
Ref.\cite{buras}. Based on the work of Ref.\cite{buras}, we have worked out the
full $10\times 10$ matrices for ${\bf \hat r}_{s,e}$ and carried out the
calculation using the full matrices.

We also need to treat the matrix elements to one-loop leve for consistency.
These one-loop matrix elements can be rewritten in terms of the tree-level
matrix elements $<O_j>^{tree}$ of the effective operators, and one
obtains\cite{fleischer,palmer}
\begin{eqnarray}
< c_i O_i> = \sum_{ij} c_i(\mu) [\delta_{ij} +{\alpha_s\over 4\pi}m^s_{ij}
+{\alpha_{em}\over 4\pi}m^e_{ij}] <O_j>^{tree}\;.
\end{eqnarray}
We have worked out the full matrices $m^{s,e}$. For the processes we are
considering only $\bar c_{3-10}$ contribute.
Expressing the effective coefficients $c^{eff}_i$ which multiply
the matrix elements
$<O_j>^{tree}$ in terms of $\bar c_i$, we have
\begin{eqnarray}
c^{eff}_3 &=&\bar c_3 - P_s/3\;,\;\; c^{eff}_4 =\bar c_3 +P_s\;,\;\;
c^{eff}_5 =\bar c_5 - P_s/3\;,\;\;c^{eff}_6 =\bar c_6 + P_s\;,\nonumber\\
c^{eff}_7 &=&\bar c_7 +P_e\;,\;\;\;\;\;\;c_8^{eff} = \bar
c_8\;,\;\;\;\;\;\;\;\;\;\;
c_9^{eff} = \bar c_9 +P_e\;,\;\;\;\;\; c_{10}^{eff} = \bar c_{10}\;.
\end{eqnarray}
The leading contributions to $P_{s,e}$ are given by:
 $P_s = (\alpha_s/8\pi)\bar c_2 (10/9 +G(m_c,\mu,q^2))$ and
$P_e = (\alpha_{em}/9\pi)(3\bar c_1+\bar c_2) (10/9 + G(m_c,\mu,q^2))$. Here
$m_c$ is the charm quark mass which we take to be 1.35 GeV. The function
$G(m,\mu,q^2)$ is give by
\begin{eqnarray}
G(m,\mu,q^2) = 4\int^1_0 x(1-x) \mbox{d}x \mbox{ln}{m^2-x(1-x)q^2\over
\mu^2}\;.
\end{eqnarray}
In the numerical calculation, we will use $q^2 = m_b^2/2$ which represents the
average value and the full expressions for $P_{s,e}$.

Using range of values of $\alpha_s(m_Z)$ and $m_t$ we can calculate the
coefficients at $\mu = m_b$.
We use $\alpha_s(m_Z)$ as input instead of $\Lambda_4$ as in Ref.\cite{buras}.
In Table 1, we show some sample WCs for the central world
average value of $\alpha_s(m_Z)=0.118$\cite{as}
and for several values of $m_t$ with $\alpha_{em} = 1/128$.

In the two-Higgs-doublet model, there are new contributions to
$c_i$ due to charged Higgs boson. The charged Higgs-quark couplings are given
by
\cite{hh1}
\begin{eqnarray}
L_H = {g\over 2\sqrt{2}m_W} \bar u_i V_{ij}
[cot\beta m_{u_i}(1-\gamma_5)- a m_{d_j}(1+\gamma_5)]d_j H^+
+H.C.\;,
\end{eqnarray}
where $cot\beta = v_1/v_2$; $v_1$ and $v_2$ are the vacuum expectation values
of the Higgs doublets $H_1$ and $H_2$,  which generate masses for down
and up quarks, respectively. The parameter $a$ depends on the models\cite{hh1}.
The main
contributions are from
the first term in eq.(8) and we will neglect the contribution from the second
term. The charged Higgs contributions to gluonic penguin
have been studied by several groups\cite{charge}. The leading QCD corrected
Hamiltonian has been given in Ref.\cite{burash}. We have checked the
next-to-leading initial conditions for the WCs at $m_W$. We
find that the inclusion of charged Higgs will not
change the initial conditions for $c_{1,2,8,10}$, but $c_{3,4,5,6,7,9}$ are
changed in the same way as those given in eqs.(32-38) of Ref.\cite{burash}.

\noindent
${\bf B \rightarrow X_s \phi}$

Using $H_{\Delta B = 1}$ in eq.(1), we obtain the
decay amplitude for $B\rightarrow X_s \phi$
\begin{eqnarray}
A(B \rightarrow X_s \phi) \approx A(b\rightarrow s \phi)
= - {g_\phi G_F\over \sqrt{2}}V_{tb}V^*_{ts}\epsilon^\mu
C \bar s \gamma_\mu (1-\gamma_5)b\;,
\end{eqnarray}
where $\epsilon^\mu$ is the polarization
of the $\phi$ particle; $C = c^{eff}_3+c^{eff}_4+c^{eff}_5+\xi
(c^{eff}_3+c^{eff}_4+c^{eff}_6) - (c^{eff}_7+c^{eff}_9+c^{eff}_{10}
+\xi(c^{eff}_8 +c^{eff}_9+c^{eff}_{10}))/2$ with $\xi=1/N_c$,
where $N_c$ is the number of colors.
The coupling constant $g_\phi$ is defined
by $<\phi|\bar s\gamma^\mu s|0> = i g_\phi \epsilon^\mu$. From the experimental
value for $Br(\phi \rightarrow e^+e^-)$, we obtain
$g^2_\phi = 0.0586\; GeV^4$.

The decay rate is, then, given by
\begin{eqnarray}
\Gamma(B\rightarrow X_s \phi) = {G_F^2g_\phi^2m_b^3 \over 16\pi m_\phi^2}
|V_{tb}V^*_{ts}|^2C^2\lambda_{s\phi}^{3/2}[1 + {3\over \lambda_{s\phi}}
{m_\phi^2\over m_b^2}(1-{m_\phi^2\over m_b^2} +{m_s^2\over m_b^2})]\;,
\end{eqnarray}
where $\lambda_{ij} = (1-m_j^2/m_b^2 -m_i^2/m_b^2)^2 - 4m_i^2m_j^2/m_b^4$.

We normalize the
branching ratio to the semi-leptonic decay of $B\rightarrow X_c e \bar \nu_e$.
We have
\begin{eqnarray}
Br(B\rightarrow X_s\phi) = Br(B\rightarrow X_c e \bar \nu_e)
{|V_{tb}V^*_{ts}|^2\over |V_{cb}|^2}{12\pi^2g_\phi^2 \lambda_{s\phi}^{3/2}
\over m_\phi^2m_b^2\rho\eta} C^2[ 1 + {3\over \lambda_{s\phi}}
{m_\phi^2\over m_b^2}(1-{m_\phi^2\over m_b^2} +{m_s^2\over m_b^2})]\;.
\end{eqnarray}
In the above expression, $\rho=1-8r^2+8r^6-r^8-24r^4\mbox{ln}r$ with
$r=m_c/m_b$, is
the phase factor, and $\eta$ is the QCD
correction factor in $B\rightarrow X_c e \bar \nu_e$, respectively.
We will use $\rho = 0.5$, $\eta = 0.889$\cite{eta}, and the approximation
$|V_{tb}V^*_{ts}/V_{cb}|^2 = 1$. The branching ratio $Br(B\rightarrow X_c
e\bar \nu_e)$ is measured to be $0.108$\cite{drell}.

\noindent
{\bf Exclusive decays $B\rightarrow K\phi (K^*\phi)$ and $B_s\rightarrow
\phi\phi$}

For the exclusive decays, we will use the factorization method. We have
\begin{eqnarray}
A(B\rightarrow K(K^*)\phi) &=& -{G_F\over \sqrt{2}}V_{tb}V^*_{ts} C
<K(K^*)|\bar s \gamma_\mu(1-\gamma_5)b|B><\phi|\bar s\gamma^\mu s|0>\;,
\nonumber\\
A(B_s\rightarrow \phi\phi) &=& -{G_F\over \sqrt{2}}V_{tb}V^*_{ts} C
<\phi|\bar s \gamma_\mu(1-\gamma_5)b|B_s><\phi|\bar s\gamma^\mu s|0>\;.
\end{eqnarray}

We can parametrize the matrix elements as
\begin{eqnarray}
<K|\bar s \gamma_\mu(1-\gamma_5)b|B> &=& f^+(q^2)(p^B_\mu + p^K_\mu)
+ f^-(q^2)q_\mu\;\nonumber\\
<v|\bar s \gamma_\mu(1-\gamma_5)b|B>&=& 2V(q^2)i\epsilon_{\mu\nu\lambda\sigma}
\epsilon_v^\nu p^{v\lambda}p^{B\sigma} \\
&+& A_1(q^2)(m_v^2-m_B^2)\epsilon_v^\mu
-A_2(q^2)\epsilon_v\cdot q (p^B_{\mu}+p^v_\mu)\;,\nonumber
\end{eqnarray}
where $v$ is a vector meson particle and $\epsilon_v^\mu$ its polarization. For
$B\rightarrow K\phi$, $q = p^B-p^K$, and for $B\rightarrow v\phi$,
$q = p^B-p^v$.

In terms of the form factors defined above, we obtain the decay rates
\begin{eqnarray}
\Gamma(B\rightarrow K\phi)&=&
{G_F^2f^{+2}(m_\phi^2)g_\phi^2m_B^3\over 32\pi m_\phi^2}|V_{tb}V^*_{ts}|^2
C^2\lambda^{3/2}_{K\phi}\;,\nonumber\\
\Gamma(B\rightarrow v\phi) &=&
{G_F^2g_\phi^2m_B^3\over 32\pi}|V_{tb}V^*_{ts}|^2
C^2\lambda^{3/2}_{v\phi}[2V^2(m_\phi^2)
+ {3\over \lambda_{v\phi}}
(1-{m_v^2\over m_B^2})^2A_1^2(m_\phi^2) - A^2_2(m_\phi^2) \nonumber\\
&+&{1\over 4m_v^2m_\phi^2}
((m_v^2-m_B^2)A_1(m_\phi^2) - (m_B^2-m_v^2-m_\phi^2)A_2(m_\phi^2))^2]\;.
\end{eqnarray}

To finally obtain the branching ratios,
we will use two sets of form factors obtained by Bauer et.
al.\cite{buar}
and Casalbuoni et. al.\cite{gatto2}.
Note that we have used different normalization for the form factors $V$ and
$A_i$ from those in Refs. \cite{buar,gatto2}.
The form factors at $q^2=0$ are determined by
using relativistic quark model in Ref.\cite{buar}, and by using chiral and
effective heavy quark theory in Ref.\cite{gatto2}. The form factors at $q^2=0$
 in Ref.\cite{buar}   are given by: $f^+_{K\phi} = 0.393$, $V_{K^*\phi} =
0.062\;GeV^{-1}$,
$A_{1K^*\phi} = -0.077\; GeV^{-1}$,
and $A_{2K^*\phi} = 0.056\;GeV^{-1}$. In Ref.\cite{gatto2}
the form factors at
$q^2=0$ are: $f^+_{K\phi} = 0.509$, $V_{K^*\phi} = 0.103\;GeV^{-1}$,
$A_{1K^*\phi} = -0.047\;GeV^{-1}$,
and $A_{2K^*\phi} = 0.034 \;GeV^{-1}$. In Ref.\cite{gatto2} the form factors at
$q^2=0$ for
$B_s\rightarrow \phi\phi$ are also calculated. They are:  $V_{\phi\phi} =
0.102\;GeV^{-1}$,
$A_{1\phi\phi} = -0.046\;GeV^{-1}$,
and $A_{2\phi\phi} = 0.033\;GeV^{-1}$.
In both papers, the $q^2$ dependence of all the form factors were assumed to be
of a simple pole type. We will use the pole masses used in
Refs.\cite{buar,gatto2}. It is interesting to note that the ratios between the
exclusive
decays and $B\rightarrow X_s \phi$ are independent of
the Wilson coefficients. If these ratios can be measured experimentally, they
can
test the models for the form factors. We obtain
\begin{eqnarray}
{Br(B\rightarrow K\phi)\over Br(B\rightarrow X_s \phi)}&=&
\left \{ \begin{array}{ll}
0.06\;,\;\;\;\;\;\;\;\;&\em{Ref}.\cite{buar}\\
0.1\;,\;\;\;\;\;\;\;\;\;&\em{Ref}.\cite{gatto2}\\ \end{array}
\right. \nonumber\\
\nonumber\\
{Br(B\rightarrow K^*\phi)\over Br(B\rightarrow X_s \phi)}&=&
\left \{ \begin{array}{ll}
0.057\;,\;\;\;\;\;&\em{Re}f.\cite{buar}\\
0.029\;,\;\;\;\;\;\;&\em{Re}f.\cite{gatto2}\\ \end{array} \right. \\
\nonumber\\
{Br(B\rightarrow \phi\phi)\over Br(B\rightarrow X_s \phi)}&=&
0.023\;,\;\;\;\;\;\;\;\;\;\;\em{Re}f.\cite{gatto2} \nonumber
\end{eqnarray}

We show in, Fuigure 1, the predictions for
the branching ratio $Br(B\rightarrow X_s \phi)$ in the SM as a function of top
quark mass $m_t$ and the strong coupling constant $\alpha_s(m_Z)$. The QCD
corrections turn out to be important which enhance the branching ratios by
about
30\% compared with those of without QCD corrections. There is a large
uncertainty
in the branching
ratios due to error in $\alpha_s(m_Z)$. From Figure 1, we see that the error
in $\alpha_s(m_Z)$ can induce an uncertainty of a factor 2.

The dominant contribuitons are from the gluonic penguin. There is a very small
$m_t$ dependence for the branching ratio calculated without the inclusion of
the electroweak penguin contributions. The inclusion
of the full electroweak contribuitons have sizeable effects which reduce
the branching ratios by
about 20\% to 30\% for the central value of $\alpha_s$ with $m_t$ varying
from 100 GeV to 200 GeV.
It is clear from Figure 1 that the full contribution has a large $m_t$
dependence.

There may be corrections to the branching ratios predicted
by the factorization method. It is a common practice to parameterize the
possible new contributions by treating  $\xi$ as a free
parameter\cite{buar,gatto2,desh3}. Using experimental values from non-leptonic
$B$ decays, it is found that\cite{gatto2},  $a_1=c_2 +\xi c_1$ and $a_2 =
c_1+\xi c_2$ have the same signs, and $|a_2|\approx 0.27$ and $|a_1| \approx
1.0$. We see that $\xi$ is close to 1/2. To
see the effect of varying $\xi$, we plot the predictions for the branching
ratios for $\xi = 1/2$ and $\xi = 1/3$. The branching ratios for $\xi=1/2$ are
about 2 times those for $\xi =1/3$.

For the central value of $\alpha_s(m_Z)$ and the central value of $m_t = 174$
GeV reported by CDF\cite{cdf}, the value for
$Br(B\rightarrow X_s \phi)$ is about $ 1.7\times 10^{-4}$ for $\xi = 1/2$.
The exclusive branching ratios $B\rightarrow
K\phi$ and $B\rightarrow K^*\phi$ are
about the same which are $1\times 10^{-5}$ if the form factors from
Ref.\cite{buar} are used. If the form factors from Ref.\cite{gatto2} are used,
one obtains
 $Br(B\rightarrow K\phi)\approx 1.7\times 10^{-5}$,
 $Br(B\rightarrow K^*\phi)\approx 0.5\times 10^{-5}$, and
$Br(B_s\rightarrow \phi\phi)\approx 0.4 \times 10^{-5}$.

In Figure 2, we show
the ratio of the branching ratios $Br(B\rightarrow X_s\phi)_{2H}$ and
$Br(B\rightarrow X_s\phi)_{SM}$ predicted by the two Higgs doublet model and
the SM as a function of $cot\beta$ for $m_t = 174$ GeV and different values of
$m_H$ with $\xi = 1/2$. The depence on $\xi$ is small. From Figure 2,
we see that the effects of the charged Higgs boson contributions are small for
$cot\beta < 1$. When increasing $cot\beta$, the charged
Higgs contributions become important and the effect is to cancel the SM
contributions. When $cot\beta$ becomes very large the charged Higgs boson
contributions become the dominant ones. However, using the information from
$B\rightarrow X_s \gamma$, it is found that for
small $m_H\sim 100$ GeV and $m_t \sim 174$ GeV, $cot\beta$ is constrained to
be less than 1\cite{hh2}. For these values, the charged Higgs boson effects on
the processes discussed in this paper are less than 10\%. For $m_H \sim 500$
GeV, the charged Higgs boson effects can reduce the hadronic penguin $B$
decays by 40\% because the range of $cot\beta$ allowed from $b\rightarrow
s\gamma$ is now larger\cite{hh2}. The effects become smaller for larger
$m_H$.

The analyses carried out in this letter can be generalized to other hadronic
$B$ decays. We will present the full calculations for the Wilson coefficeints,
the full expressions for $P_{s,e}$ and other related decays in a forthcoming
paper\cite{dh}.

We thank Buras, McKellar, Fleischer for useful corespondences and
thank Lautenbacher for many useful discussions.

\newpage

\begin{table}
\caption{The Wilson coefficients for $\Delta B = 1$ at $m_b = 5\; GeV$ with
$\alpha_s(m_Z) = 0.118$.}
\begin{tabular}{|c|c|c|c|c|c|c|c|c|c|c|}
$m_t$(GeV) & $\bar c_1$ &$\bar c_2$ &$\bar c_3$ &$\bar
c_4$&$\bar c_5$&$\bar c_6$&$\bar c_7/\alpha_{em}$&$\bar c_8/\alpha_{em}$&$\bar
c_9/\alpha_{em}$&$\bar c_{10}/\alpha_{em}$\\ \hline
130&-0.313&1.150&0.017&-0.037&0.010&-0.045&-0.061&0.029&-0.978&0.191\\ \hline
174&-0.313&1.150&0.017&-0.037&0.010&-0.046&-0.001&0.049&-1.321&0.267\\ \hline
210&-0.312&1.150&0.018&-0.038&0.010&-0.046&0.060 &0.069&-1.626&0.334
\end{tabular}
\label{table1}
\end{table}
\newpage

\begin{center}
Figure Captions
\end{center}

Figure 1. $Br(B\rightarrow X_s \phi)$ as a function of $m_t$ and $\alpha_s
(m_Z)$. The regions between the dashed and solid lines are the branching
ratios for $\alpha_s(m_Z)$ varying from 0.111 to 0.125 for $\xi = 1/2$ and
$\xi=1/3$, respectively. The branching ratios increases with $\alpha_s(m_Z)$.

Figure 2. $Br(B\rightarrow X_s\phi)_{2H}/Br(B\rightarrow X_s \phi)_{SM}$ as
a function of $cot\beta$, and $m_H$. The curves 1, 2 and 3 are for
$m_H$ equals to 100, 500 and 1000 GeV, respectively.


\begin{references}
\bibitem{bdecay} For a review see: $B\; Decays$, edited by S. Stone, World
Scientific, 1992.
\bibitem{cleo} CLEO Collaboration, Phys. Rev. Lett. {\bf 71}, 674(1993).
\bibitem{desh1} N.G. Deshpande et. al., Phys. Rev. Lett. {\bf 59}, 183(1987);
S. Bertolini, F. Borzumati and A. Masiero, $ibid$, 180(1987);
N.G. Deshpande, P. Lo and J. Trampetic, Z. Phys.{\bf C40},
369(1988); C. Dominguez, N. Paver and Riazuddin, Phys. Lett. {\bf B214},
459(1988); A. Ovchinnikov and V. Slobodenyuk, Phys. Lett. {\bf B237},
569(1990); P.J. O'Donnel and H.K.K. Tung, Phys. Rev.{\bf D44}, 741(1991);
R. Casalbuoni et. al., Phys. Lett. {\bf B312}, 315(1993).
\bibitem{desh2}N.G. Deshpande and J. Trampetic, Phys. Rev. {\bf D41},
895(1990).
\bibitem{gatto1}A. Deandrea, et. al., Phys. Lett. {\bf B320}, 170(1993).
\bibitem{fleischer} R. Fleischer, Preprint, TUM-T31-40/93 (Z. Phys. in press).
\bibitem{laut} A. Buras, M. Jamin, M. Lautenbacher and P. Weisz, Nucl. Phys.
{\bf B400}, 37(1993); A. Buras, M. Jamin and M. Lautenbacher, ibid, 75(1993);
M. Ciuchini, E. Franco, G. Martinelli and L. Reina, Nucl. Phys. {\bf B415},
403(1994).
\bibitem{bruce} F. Gilman and M. Wise, Phys. Rev. {\bf D20}, 2392(1979);
R. Miller and B. McKellar, Phys. Rept. {\bf 106}, 169(1984).
\bibitem{buras} A. Buras, M. Jamin, M. Lautenbacher and P. Weisz,
Nucl. Phys. {\bf B370}, 69(1992).
\bibitem{palmer} R. Fleischer, Z. Phys. {\bf C58}, 483(1993);
G. Kramer, W. Palmer and H. Simma, Preprint, DESY-93-192.
\bibitem{as} S. Bethke, in Proceedings of the XXV International Conference on
High Energy Physics, Dallas, Texas, August, 1992.
\bibitem{hh1} J. Gunion, H. Haber, G. Kane and S. Dawson, $The\; Higgs\;
 Hunter's\; Guide$ (Addison-Wesley, Redwood City, CA 1990).
\bibitem{charge} Wei-Shu Hou and R.S. Willey, Phys. Lett. {\bf B202},591(1988);
Wei-Shu Hou, Nucl. Phys. {\bf B326}, 54(1989);  V. Barger, J. Hewett and
R. Phillips, Phys. Rev. {\bf D41}, 3421(1990); A. Davies, G.C. Joshi and M.
Matsuda, Z. Phys. {\bf C52},97(1991); A. Davies, T. Hayashi, M. Matsuda and
M. Tanimoto, Preprint, AUE-02-93.
\bibitem{burash} G. Buchalla, A. Buras, M. Harlander, M. Lautenbacher and
C. Salazar, Nucl. Phys. {\bf B355}, 305(1991).
\bibitem{eta}N. Cabbibo and L. Maiani, Phys. Lett. {\bf B79}, 109(1978);
M. Suzuki, Nucl. Phys. {\bf B145}, 420(1978); N. Cabbibo, G. Corbe and
L. Maiani, Phys. Lett. {\bf B155}, 93(1979).
\bibitem{drell}P. Drell, in Proceedings of the XXV International Conference on
High Energy Physics, Dallas, Texas, August, 1992.
\bibitem{buar}M. Bauer, B. Stech and M. Wirbel, Z. Phys. {\bf C34}, 103(1087).
\bibitem{gatto2}A. Deandrea, N. Di Bartolomeo, R. Gatto and G. Nardulli,
Phys. Lett. {\bf B318}, 549(1993).
\bibitem{desh3} N.G. Deshpande, M. Gronau and D. Sutherland, Phys. Lett.
{\bf B90}, 431(1980).
\bibitem{cdf} F. Abe, et al., CDF Collaboration, Preprint,
FERMILAB-PUB-94/097-E, CDF/PUB /TOP/PUBLIC/2561.
\bibitem{hh2} J.L. Hewett, Phys. Rev. Lett. {\bf 70}, 1045(1993); V. Barger,
M. Berger, R. Phillips, $ibid$, 1368(1993).
\bibitem{dh} N. Deshpande and Xiao-Gang He, in preparation.

\end{references}
\end{document}